# DeepAdjoint: An All-in-One Photonic Inverse Design Framework Integrating Data-Driven Machine Learning with Optimization Algorithms

*Christopher Yeung[1,2], Benjamin Pham[1], Ryan Tsai[1], Katherine T. Fountaine[2], and Aaswath P. Raman[1,*]*

[1]Department of Materials Science and Engineering, University of California, Los Angeles, CA 90095, USA
[2]NG Next, Northrop Grumman Corporation, Redondo Beach, CA 90278, USA
*Corresponding Author: aaswath@ucla.edu

**ABSTRACT:** In recent years, hybrid design strategies combining machine learning (ML) with electromagnetic optimization algorithms have emerged as a new paradigm for the inverse design of photonic structures and devices. While a trained, data-driven neural network can rapidly identify solutions near the global optimum with a given dataset's design space, an iterative optimization algorithm can further refine the solution and overcome dataset limitations. Furthermore, such hybrid ML-optimization methodologies can reduce computational costs and expedite the discovery of novel electromagnetic components. However, existing hybrid ML-optimization methods have yet to optimize across both materials and geometries in a single integrated and user-friendly environment. In addition, due to the challenge of acquiring large datasets for ML, as well as the exponential growth of isolated models being trained for photonics design, there is a need to standardize the ML-optimization workflow while making the pre-trained models easily accessible. Motivated by these challenges, here we introduce DeepAdjoint, a general-purpose, open-source, and multi-objective "all-in-one" global photonics inverse design application framework which integrates pre-trained deep generative networks with state-of-the-art electromagnetic optimization algorithms such as the adjoint variables method. DeepAdjoint allows a designer to specify an arbitrary optical design target, then obtain a photonic structure that is robust to fabrication tolerances and possesses the desired optical properties – all within a single user-guided application interface. We demonstrate DeepAdjoint for the design of infrared-controlled metasurfaces, and show that a wide range of structures and absorption spectra can be achieved and optimized, including single- and multi-resonance behavior through single- and supercell-class structures, respectively. Our framework thus paves a path towards the systematic unification of ML and optimization algorithms for photonic inverse design.

# Introduction

Photonic structures and devices are now an essential component of a broad range of information, life sciences, and renewable energy technologies. Some examples include plasmonic waveguides for photonic integrated circuits[1,2], optical filters for spectroscopy and super-resolution imaging[3,4], and metasurfaces or metamaterials for flat optical components and solar energy harvesting[5,6]. However, rising demands in nanophotonic device performance and functionality have resulted in the design process becoming increasingly complex and computationally intensive[7]. For instance, subwavelength dielectric and metallic nanostructured materials can be structured into complex geometric configurations that scatter, localize, and/or tailor electromagnetic fields to achieve new modalities in light-matter interactions[8]. Due to the wide choice of materials and the spatial degrees-of-freedom available for design, the design space one must explore in contemporary photonic inverse design is typically highly nonlinear and non-convex (*i.e.*, contains many local optima), and thus extremely challenging and time-consuming to navigate.

Motivated by this challenge, machine learning (ML) and deep learning methods based on neural networks have shown tremendous promise towards addressing conventional limitations on photonic inverse design. Neural networks are capable of capturing, interpolating, and optimizing nonlinear data-based and physics-based relationships, including those found in nanophotonic systems. Neural networks achieve such capabilities by building an implicit relationship between input and output responses, which for nanophotonics inverse design are the optical responses and geometric/material parameters, respectively. A trained neural network is orders of magnitude faster than typical full-wave simulations and can generate non-intuitive physical structures in response to desired optical properties[9]. Accordingly, a substantial number of studies have employed neural networks for designing a broad range of photonic systems, including: metasurfaces[10,11,36,43-45], photonic crystals[12,13], and plasmonic nanostructures[14,15]. However, despite numerous advancements, it is well-known that neural networks cannot generalize too far beyond the information available in the training dataset[16-18]. Due to these limitations, *hybrid* algorithms combining deep learning and conventional optimization methods have emerged as a new class of efficient inverse design methodology[19,20].

Recent studies have integrated different types of neural networks and optimization schemes for photonic inverse design. Early works paired neural networks with particle swarm optimization[21] (PSO) and evolutionary algorithms[22,23]. Emerging works also have integrated deep learning with optimization through more sophisticated design pipelines[41-45]. Collectively, these studies successfully showed that the neural network can perform a rough estimate of the desired solution (*i.e.*, a global search), while the iterative optimization algorithm carries out an additional refinement step (*i.e.*, a local search). Since conventional optimization algorithms need an ideal initial condition in order to obtain the optimal result, and the neural network is restricted by its training data, the combination of both techniques can simultaneously overcome their individual limitations[20,24]. Recent hybrid ML-optimization approaches have also employed more advanced neural networks and optimization algorithms. For example, generative adversarial networks[25] (GANs) and variational autoencoders[26] (VAEs) were used together with the adjoint variables method for photonic design. Adjoint-based optimization is one of the most widely-used algorithms for photonics inverse design because regardless of the number of elements in the design space, the algorithm can determine the shape or topology gradient using only a forward and adjoint (time-reversed) simulation at each iteration[27-29]. As a result, GANs and VAEs can design complex topological structures through image-based representations, while the adjoint method can efficiently push performance further[30]. Additionally, generative models trained on physics-informed losses (or using the adjoint method within the training process) have also benefited from a subsequent optimization-based refinement step[31]. Thus, a number of deep learning models have been trained across various photonic device types, and a precedent has been established for hybrid ML-optimization algorithms as the next generation of inverse design methods. However, a hybrid ML-optimization strategy that simultaneously optimizes across multiple materials and geometries has yet to be realized. Moreover, the integration between a data-driven ML model and optimization algorithms typically involve elaborate and highly-specialized procedures. For example, to establish a link between neural networks and conventional optimization methods, intermediate steps are required to introduce robustness, convert file formats, and/or to ensure that the network outputs can be adapted to the algorithm of interest.

To streamline the ML-optimization process, here we introduce an "all-in-one" global inverse design application framework which seamlessly combines generative networks with adjoint-based optimization algorithms to simultaneously optimize across materials and geometries. "Global" in this context refers to the network's ability to perform a global search within the surveyed design space, which includes material properties and freeform topology, but the network does not guarantee that the final generated device is globally optimal. Schematically illustrated in Figure 1, our framework, DeepAdjoint, allows a researcher to specify an arbitrary spectral target (labeled "1" in Figure 1) and pass the target directly into a pre-trained generative network. Such pre-trained models can be data-driven[38] (*i.e.*, trained on loss functions that quantify the error between training data and model predictions), physics-driven[39] (*i.e.*, trained on partial differential equations that capture governing physical laws), or a combination thereof[40]. In this regard, we note that the increasing number of deep learning models being generated for photonics design (which we expect will continue to grow exponentially in the near future) reinforces the need for a design process that can integrate pre-trained models, particularly when practices such as network sharing and model serving are expanding within the machine learning community[32,34].

As a proof of concept, we employed a global inverse design GAN model with the ability to simultaneously predict device class, material properties (*e.g.;* refractive index and Drude plasma frequency), and nanoscale geometric structuring (including planar topology and layer thickness) for metal-insulator-metal (MIM) metasurfaces[9]. After passing the target into the GAN (labeled "2" in Figure 1), DeepAdjoint then validates the GAN-generated design using full-wave numerical simulations (labeled "3" in Figure 1). As a default simulation tool, DeepAdjoint integrates directly with a commercial finite-difference time-domain solver (Lumerical FDTD). The GAN-generated design can then be further augmented by converting the design into an adjoint optimization procedure (labeled "4" in Figure 1), after which the final design can yield even greater accuracy or performance by extending beyond the model's limitations (labeled "5" in Figure 1). We demonstrate this end-to-end workflow for a range of optical device targets, including single- and multi-resonance responses, for infrared-controlled MIM metasurfaces.

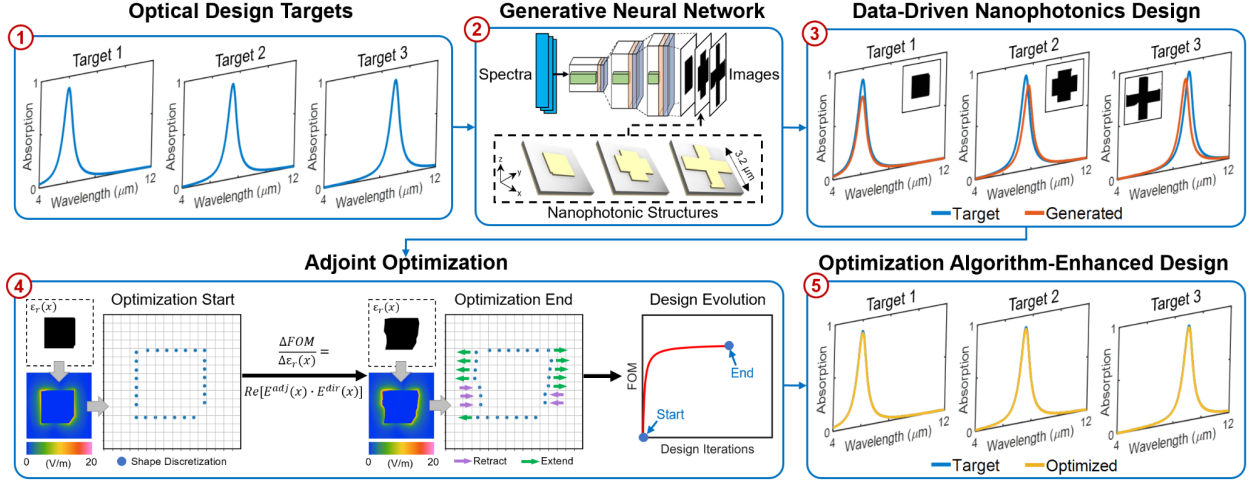

**Figure 1.** DeepAdjoint: an photonics inverse design framework schematic combining deep learning and adjoint optimization. (1) An arbitrary optical design target can be specified and (2) passed into a pre-trained neural network to generate a nanophotonic structure that is (3) validated using full-wave FDTD simulations. (4) The network's design can then be automatically converted into an adjoint optimization procedure, which can (5) yield device accuracy or performance that extends beyond the network's potential limitations.

## Methods

To democratize the hybridization of deep learning with electromagnetic optimization, and to make our framework easily-accessible to a wide range of practitioners, we deployed and packaged DeepAdjoint as a standalone application with a user-guided interface. Figure 2 presents the details of the application, where each step in Figure 1 can be executed (and the results can be observed) within a single user-friendly environment. As an example step-by-step procedure for designing MIM metasurfaces, DeepAdjoint first allows the user to specify an input target absorption spectrum (labeled "1" in Figure 2). Here, a Lorentzian function with a center wavelength of 6 μm and full width half maximum (FWHM) of 0.75 μm is defined and shown within the built-in visualization tool (blue curve). Next, the user simply imports the generative model, then generates the design (in ~500 ms) with a single button press at the step labeled "2" in Figure 2. In our implementation of DeepAdjoint, we leveraged a conditional deep convolutional generative adversarial network (DCGAN) that was developed within a prior study[9], which facilitates the simultaneous prediction of the material properties (*i.e.*, refractive index and plasma frequency), layer thicknesses, and planar geometries of photonic structure. This particular model was trained on 20,000 metasurface designs (of which 10% was reserved for validation) derived from various shape templates: cross, square, ellipse, bow-tie, etc. The model inputs were 800-point absorption spectrum vectors, while the model outputs were MIM metasurface designs (in 3.2 × 3.2 μm² unit cells) represented as 64 × 64 × 3 pixel "RGB" images. Details of the optimized model architecture can be found in the Supporting Information.

Since the direct output of the GAN is a set of matrix values and must be converted into a simulation model for numerical analysis, with the press of another button, DeepAdjoint converts the GAN's output into an FDTD model of the metasurface, runs the simulation, then reports the results back into the user interface for comparison (labeled "3" in Figure 2). Following this step, the FDTD-validated absorption spectrum (orange curve) and corresponding electric field profiles can be observed directly on the application interface.

Next, the GAN's design can be enhanced by applying the adjoint optimization method (labeled "4" in Figure 2), where an optimization target wavelength can be specified that the algorithm aims to maximize. To execute the adjoint optimization procedure, we implemented a customized version of the LumOpt module[37] (a Python wrapper for Lumerical FDTD). In this particular implementation, a number of enhancements were made to the base module in order to support free-space reflective metasurface design and optimization, which we summarized in Figure S3 of the Supporting Information. At the time the study was conducted, the adjoint optimization module we employed was limited to only the optimization of a photonic structure's planar geometry, and thus the material properties and layer thicknesses (derived from the GAN's predictions) remained fixed during optimization. We note that our demonstration of DeepAdjoint also leveraged a deep learning model trained exclusively on polarization-dependent designs. Accordingly, the proceeding adjoint-optimized structures were optimized specifically for single-polarization performance at normal incidence. However, the presented methodology is generalizable to polarization-independent structures and optimization beyond just the planar geometries if, for instance, the integrated model was trained with the corresponding designs and the optimization module supported multi-dimensional exploration, respectively.

To configure the GAN's design for adjoint optimization, an automatic multistep process is performed (shown in Figure S1 of the Supporting Information), where the GAN's output is refined (by removing voids and defects) and transformed into a set of discretized polygon points at the meta-atom or resonator boundary. In doing so, the polygon points (*i.e.*, optimizable parameters) are compatible with the adjoint shape optimization process. Then, as the adjoint optimization iteratively progresses, the coordinates of the polygon points gradually change in the direction of figure-of-merit (FOM) improvements (labeled "5" in Figure 2). Moreover, since the presented metasurface designs operate in a reflective manner at normal incidence, the typical forward and adjoint simulations required are identical here and can be reduced to a single simulation at each iteration. Thus, we note that our particular implementation of the adjoint optimization algorithm has increased computational efficiency for metasurface design. Additionally, our framework allows the user to specify minimum feature sizes and fabrication tolerances without sacrificing device performance (shown in Figure S2 of the Supporting Information).

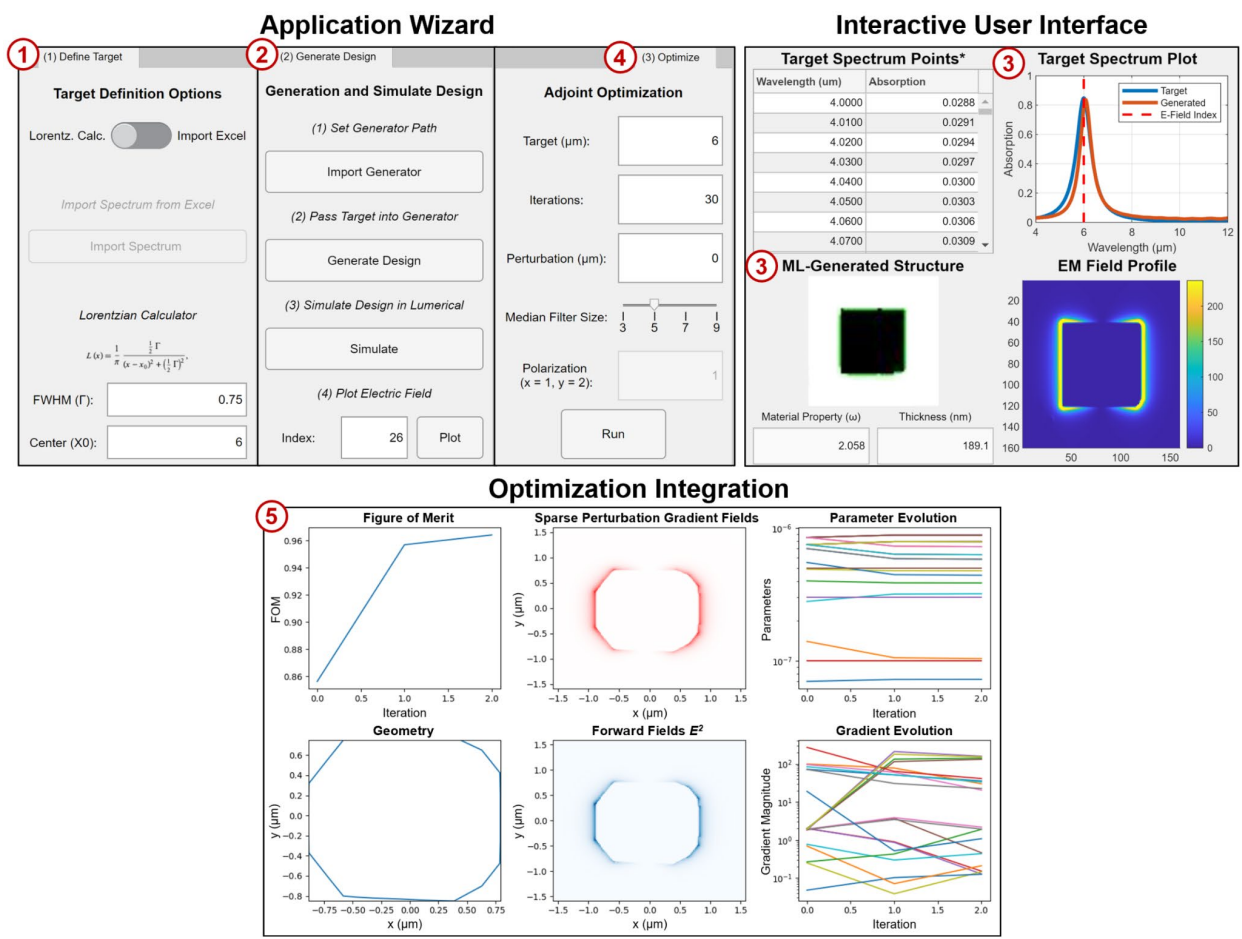

**Figure 2.** DeepAdjoint application interface and step-by-step workflow. Users can (1) define targets, (2) generate designs, (3) validate designs, (4) run adjoint optimizations, and (5) monitor optimization results.

# Results and Discussion

To highlight the advantages of the proposed framework, we first investigated the performance of the adjoint optimization algorithm in relation to the algorithm's initial designs (for the particular MIM structure design space we evaluated in this work). In Figure 3, three adjoint optimization runs were executed at three different target wavelengths (indicated by the dashed red lines): 6, 7, and 8 μm, which are presented in Figures 3a, 3b, and 3c, respectively. Each optimization was performed using randomized starting designs (orange lines), and the objective was to maximize horizontal polarization ($\theta$=0) absorption at the designated target wavelength. At the end of the optimization runs, we observe that the final designs (yellow lines) typically exhibited higher absorption values/peaks than the initial designs. Center and right columns of Figure 3 show symmetric and asymmetric starting designs, respectively. Importantly, we note that different starting designs yielded different degrees of performance improvements (*i.e.*, different absorption peak amplitudes). Moreover, it can be observed that several optimized designs possess extra absorption peaks (beyond the target wavelengths) that were originally unintended. In several instances, as shown in the left column of Figure 3, a poor starting design can also cause the adjoint optimization to fail by not finding noticeable improvements to the initial structure. In Figure S4 of the Supporting Information, we further assess the general boundaries of this failure phenomenon, which is explained by the starting design being too far from the sought target. Thus, a deep learning algorithm that can provide the optimization with an ideal starting design would not only save computation time by reducing the number of optimization iterations, but also allow the optimization to succeed and reach an optimal solution without any excess optical behaviors.

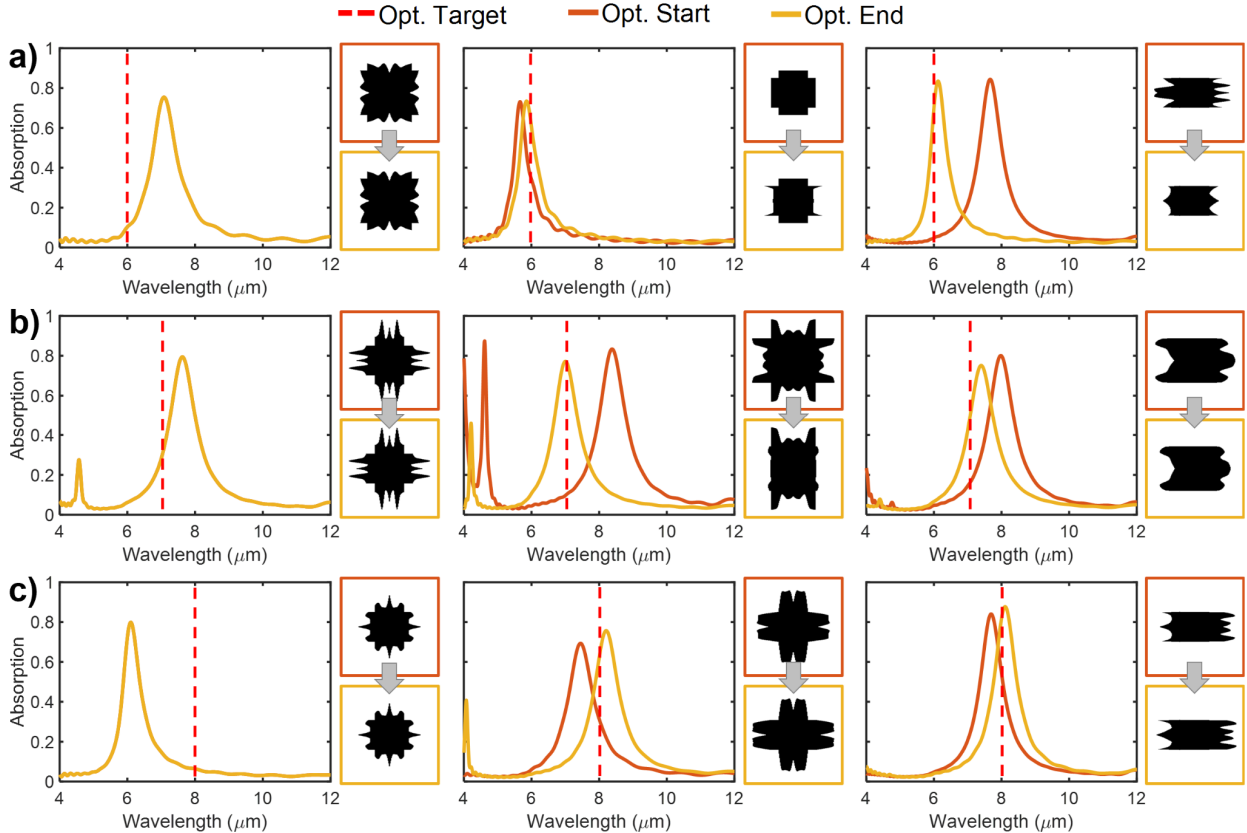

**Figure 3.** Metasurface designs created via adjoint optimization with randomized initial designs (orange lines). Optimization objectives include maximizing horizontal polarization ($\theta$=0) absorption at (a) 6, (b) 7, and (c) 8 µm target wavelengths (red dashed lines). Optimized structures and corresponding absorption spectra (yellow lines) possess various degrees of performance improvements and several extra unintended absorption peaks due to the random, suboptimal nature of the starting designs. The left column shows instances where the adjoint optimization fails to noticeably improve the starting design. Center and right columns show adjoint optimization results with symmetric and asymmetric starting designs, respectively.

Next, we applied our DeepAdjoint framework to the optimization of metasurfaces with single-resonance absorption peaks. Figure 4 presents a series of optimized designs, generated through DeepAdjoint, using a range of input absorption spectra (blue lines) with "hand-drawn" Lorentzian-shaped peaks from 5 to 9 µm. Here, we observe that the GAN's designs and simulated spectra (orange lines) are close matches to the input targets. However, several designs possess off-centered peaks or lower amplitudes in comparison to the original target. After using the GAN-generated designs as the starting points for subsequent adjoint optimization runs (with the optimization targets marked by the red dashed lines), it can be observed that the off-centered peaks are rectified and the low-amplitude peaks are increased by up to 75% (compared to the starting spectra). Moreover, the final absorption peaks of DeepAdjoint's designs are 10% higher than the best adjoint optimization-only designs using random starting points (from Figure 3).

In Figures S4 and S5 of the Supporting Information, we evaluated the computation times between traditional adjoint optimization with random starting designs and DeepAdjoint, respectively. From these results, we observed that DeepAdjoint reduced optimization iterations by more than 50%. On a distributed high-performance computing cluster with four dedicated compute nodes, where a node has a minimum of four 64-bit Intel Xeon or AMD Opteron CPU cores and 8 GB memory, the DeepAdjoint optimizations corresponded to approximately 1 hour of total computation time. Thus, for conditional photonics inverse design with a wide range of input targets, we demonstrate that the hybridization of generative networks with the adjoint optimization algorithm offers a number of advantages, including: superior device performance in comparison to each standalone method, increased computational efficiency, and eliminating reliance on randomized starting designs.

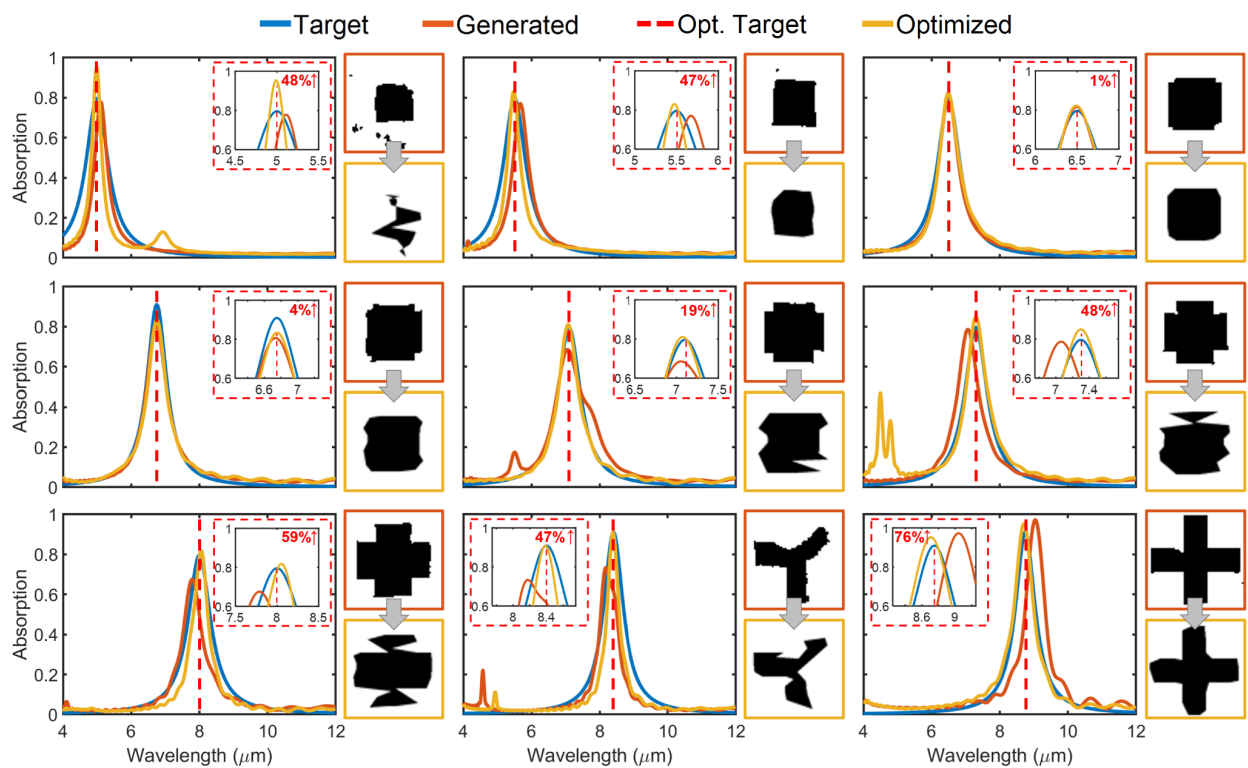

**Figure 4.** Single-objective metasurface designs (one absorption peak) created via DeepAdjoint. Target absorption spectra (blue lines) are passed into the generative model (GAN) to produce starting designs (orange lines) for adjoint optimization (red dashed lines). Optimized designs (yellow lines) exhibit up to 75% performance enhancements in comparison to GAN-generated designs (shown in the inset images) and 10% improvement over adjoint optimization-only designs

with random starting points, indicating the hybrid approach exceeds the performance of each individual method.

Because meta-structures with simple, single-resonator periodic unit cells may only offer limited capabilities[23], we next demonstrate the versatility of our ML-optimization framework by applying it to multi-objective supercell designs, where the goal is to design compound meta-atoms with multiple resonant behaviors. We note that designing such supercell structures is particularly challenging using conventional approaches, since adjacent elements may exhibit coupling and interference[35]. Furthermore, the increased number of parameters in the supercell naturally results in additional optimization complexity and computational costs. Accordingly, using DeepAdjoint, we address these challenges by first specifying the individual target resonance peaks within the supercell structure (as shown in the blue lines of Figure 5). This in turn generates the individual unit cells which contribute to the target absorption peaks (as previously demonstrated). When the individual unit cells are merged into supercell structures, it can be observed that the final structures (orange lines) produce fairly close matches in comparison to the input targets. However, compared to the single unit cell designs, the supercells have lower absorption peaks as a result of cross-element coupling. Thus, designing a supercell is not as simple as generating and combining the individual components, though this can provide a decent approximation. In this regard, a multi-objective adjoint optimization procedure can be applied to the generated supercell structures, which simultaneously maximizes multiple absorption peaks while accounting for the optical behaviors produced by the entire supercell (including cross-element coupling).

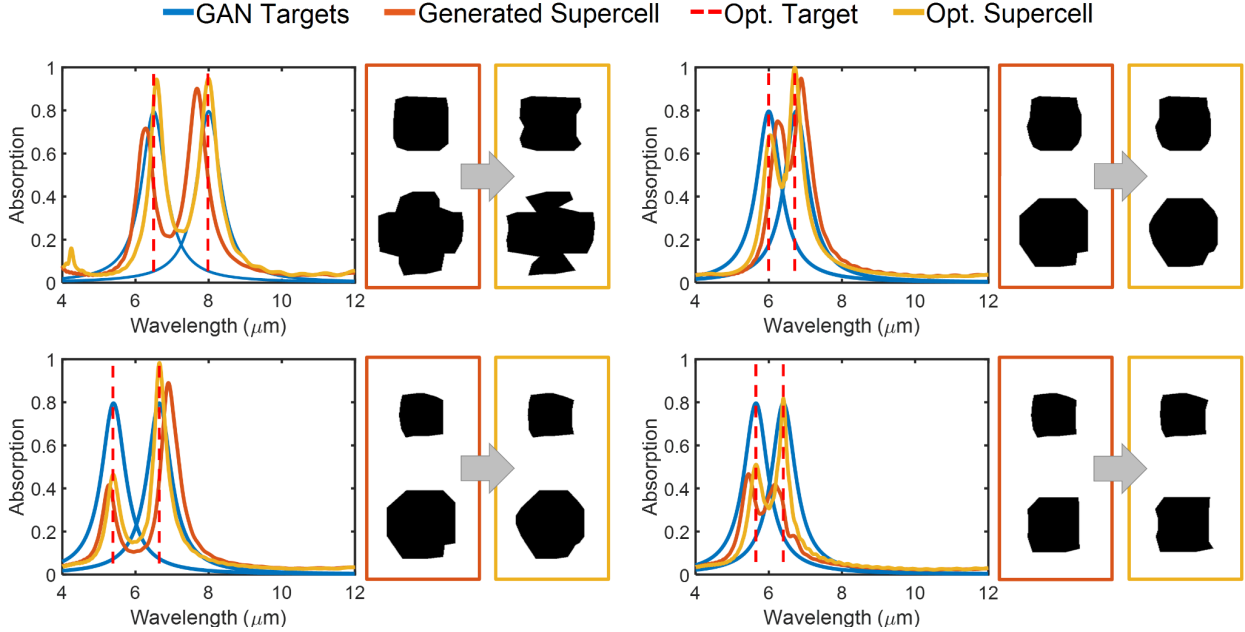

**Figure 5.** Multi-objective metasurface designs (multiple absorption peaks) created via DeepAdjoint. Target absorption spectra (blue lines) are passed into the generative model (GAN) to produce starting designs (orange lines) for adjoint optimization (red dashed lines). Optimized designs (yellow lines) exhibit up to 50% performance enhancements in comparison to GAN-generated supercell designs (shown in the inset images).

In Figure 5, the results of multiple supercell optimization runs are presented (at the optimization targets indicated by the red dashed lines). Here, we observe that the optimized supercells (yellow lines) all yield up to 50% higher absorption peaks than the initial designs, though the degree of absorption enhancement appears to be peak-dependent (possibly due to different coupling mechanisms induced by particular elements). In addition to increasing the target absorption peaks, Figure 5 also shows that the adjoint optimization procedure can rectify or recenter off-target peaks within the supercell. In Figure S6 of the Supporting Information, we validated our approach even further through the optimization of a larger (four unit cell) supercell structure. As a result, we show that our hybrid ML-optimization framework can be used to design and achieve a wide range of optical behaviors, including periodic unit cell structures with single resonances and complex supercell structures with multiple resonances or broadband characteristics. Furthermore, we note that at the core of our methodology, a pre-trained model with good performance is a key prerequisite. Therefore, a potential limitation of our approach is if the sought target is within a regime that is poorly represented by the training dataset, which would contribute to the generation of poor starting designs for optimization. To overcome these limitations, larger and more diverse training datasets may be constructed. To a further extent, in future works, automated feedback loops may also be implemented which identify areas of weak training data coverage, compensate the model with new data, then retrain the model to improve design performance and generalization.

# Conclusion

In summary, we presented DeepAdjoint, a general-purpose, open-source, and multi-objective "all-in-one" global photonics inverse design application framework that streamlines and augments the ML-optimization pipeline by integrating data-driven deep generative network with state-of-the-art electromagnetic optimization algorithms. DeepAdjoint allows a designer to specify an arbitrary optical design target, then obtain a photonic structure that is robust to fabrication tolerances and possesses the sought optical properties – all within a single user-guided workflow and application interface. As a proof of concept, we demonstrated our framework for the design and optimization of infrared-controlled metasurfaces, and showed that a wide range of structures and absorption spectra can be achieved, including single- and multi-resonance behavior through single- and supercell-class structures, respectively. By specifying an input target spectrum, a global inverse design generative neural network serves as a rapid global approximation search step (~500 ms) and produces a nanophotonic structure with material properties, layer thicknesses, and planar geometry defined. Afterwards, the generated design can be sent through an adjoint optimization procedure, which serves as a local search step to increase performance further. As a result, the limitations of training data restriction and starting point dependency for deep learning and conventional optimization respectively, can be simultaneously overcome. Our proposed framework is thus an important step towards leveraging the strengths of both data-driven machine learning and optimization algorithms for a universal photonics inverse design framework.

# Supporting Information

Generative adversarial network (GAN) to adjoint optimization configuration, minimum feature size analysis, and adjoint optimization implementation details and analyses.

# Acknowledgements

The authors would like to thank S. Larouche, A. Howes, and M. W. Knight for their meaningful discussions and contributions on the application of DeepAdjoint. This work used computational and storage services associated with the Hoffman2 Shared Cluster provided by UCLA Institute for Digital Research and Education's Research Technology Group. The code used to generate the results reported in this paper will be made available on GitHub (https://github.com/Raman-Lab-UCLA) upon publication of this manuscript and is also available from the corresponding author upon request.

# Funding Sources

This work was supported by the Sloan Research Fellowship from the Alfred P. Sloan Foundation and the DARPA Young Faculty Award (#W911NF2110345).

# Conflicts of Interest

The authors declare no conflicts of interest.

# Supporting Information

**Deep Convolutional Generative Adversarial Network (DCGAN) Architecture**

Implemented in the PyTorch framework, the DCGAN consists of two networks: a generator and a discriminator. The optimized generator contains five transposed convolutional layers (with 1200, 1024, 512, 256, and 128 input channels or feature maps), while the discriminator has five convolutional layers (with 6, 64, 128, 256, and 512 input channels or feature maps). Each transposed convolutional layer in the generator is followed by a batch normalization and ReLU (rectified linear unit) activation layer, instead of the final layer, where a Tanh (hyperbolic tangent) activation is used. Similarly, in the discriminator, each convolutional layer is followed by a batch normalization and Leaky ReLU layer, and the final layer possesses a Sigmoid activation. At the generator input layer, the 800-point absorption spectra are concatenated with 400-point latent vectors to yield 1200-point input vectors. For the discriminator input, the absorption spectra are passed through a fully-connected layer and reshaped into a 64×64×3 matrix. These matrices were then concatenated with the real and generated images to form 64×64×6 inputs for the discriminator. Model training was performed on an NVIDIA Titan RTX GPU and took approximately 30 minutes to complete.

**Generative Adversarial Network (GAN) to Adjoint Optimization Configuration**

In this work, we show that the GAN's design can be enhanced through the application of the adjoint variables method. To easily-support the adjoint variables method for shape optimization, which requires the calculation of a structure's shape derivative, the shape can be described by a 2D array of polygon points that are sorted in a counterclockwise fashion. Accordingly, we implemented an automatic multistep process (shown in Figure S1) where the GAN's image-formatted output data is refined and transformed into a set of discretized polygon points. In this process, a median filter is first applied to the GAN's output shape to remove voids, defects, and image artifacts. Then, the Canny edge detection algorithm is used to extract the meta-atom or resonator boundary. Afterwards, Hough line transformation is applied to the extracted edge in order to convert the resonator boundary into line coordinates, which we then interpolate to produce the required polygon points for shape optimization. We note that the interpolation spacing between the polygon points can be user-specified (and is 100 nm by default), which determines the resolution of the adjoint-optimized nanostructure.

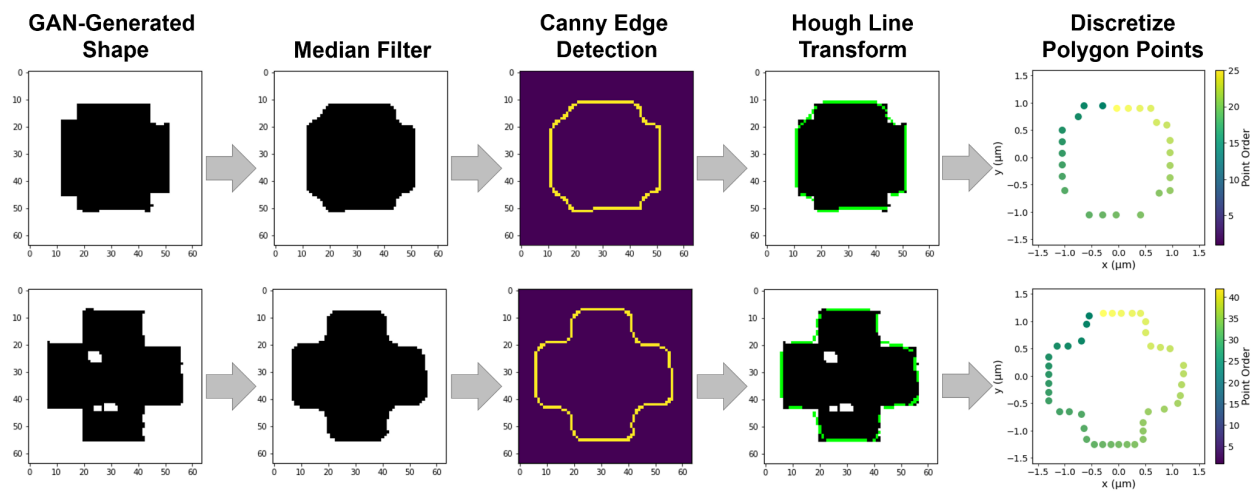

**Figure S1.** GAN to adjoint optimization configuration procedure. The GAN's output image is filtered, then passed through the Canny edge detection and Hough transform algorithms, to produce a set of discretized polygon points that are amenable to adjoint shape optimization.

**Minimum Feature Size Analysis**

During the GAN to adjoint optimization configuration process, the median filter can serve as a means to remove image artifacts and to specify minimum feature sizes. In our implementation of DeepAdjoint, we set the size of the filter as a user-defined value, such that a user can enhance the fabricability of the ML-optimization generated designs. In Figure S2, we show that the minimum feature size of 50 nm can be increased to 250 nm without significantly affecting the absorption spectra of the GAN-generated designs. As a result, designers can ensure their designs are robust to fabrication tolerances without having to retrain the entire model, although retraining a model using larger feature sizes is also a valid option that is supported by the model-loading capabilities of DeepAdjoint.

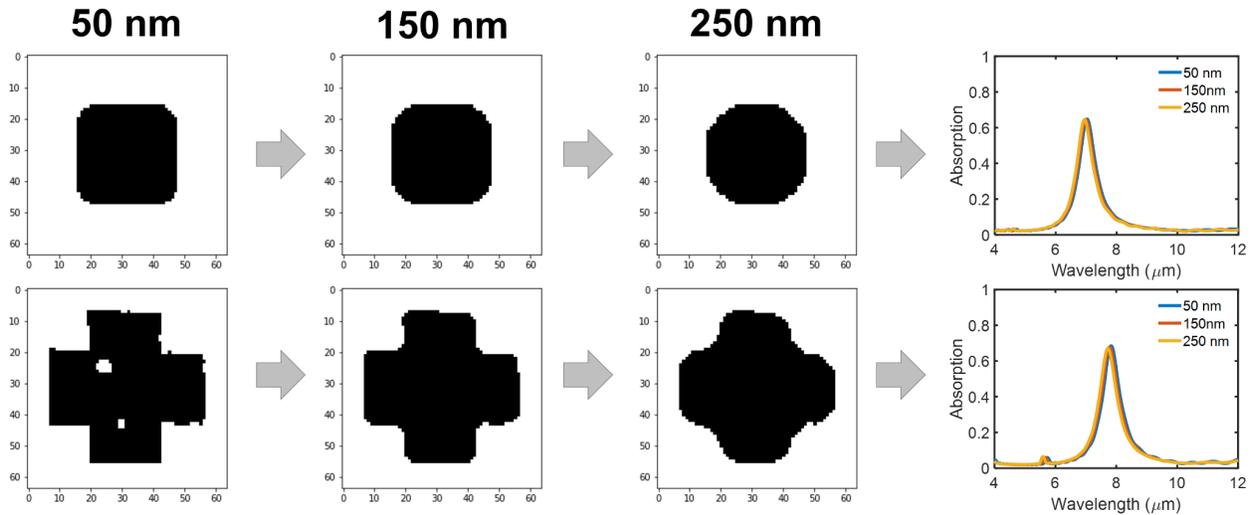

**Figure S2.** Effect of changing the median filter or minimum feature size. The minimum feature size (50 nm default) can be set to 250 nm without significantly affecting the absorption spectra.

## Adjoint Optimization Implementation Details and Features

In our particular implementation of the adjoint method, we employed a customized version of a widely-used continuous adjoint optimization Python wrapper for FDTD simulations (LumOpt; referenced in the main manuscript). Here, we note several key enhancements and modifications that were made to the base LumOpt module (which was primarily designed for integrated photonics) in order to enable free-space metasurface design and optimization. First, as shown in Figure S3a, we tailored the simulation setup and boundary conditions by applying a plane wave source instead of a mode source. In addition, since the presented metasurface designs operate in a reflective manner at normal incidence, the typical forward and adjoint simulations required are identical and were reduced to a single simulation at each iteration. Thus, we note that our particular implementation of the adjoint optimization algorithm has increased computational efficiency for metasurface design. Symmetric boundary conditions across the x- and y-planes (one plane for two-fold symmetry or two planes for four-fold symmetric) are also supported to further reduce simulation time.

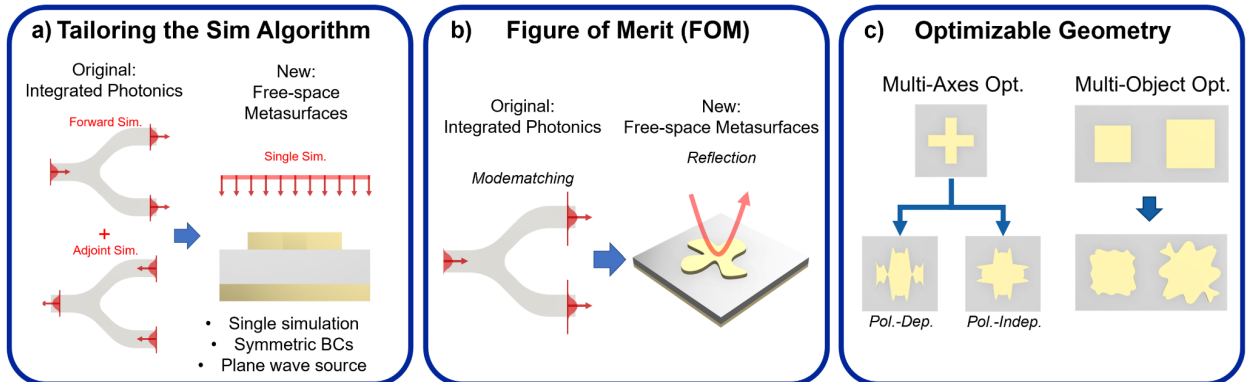

**Figure S3.** Custom implementation of the adjoint method based on numerous enhancements and modifications to the LumOpt module. Enhancements include: (a) changes to the simulation setup and boundary conditions to support free-space metasurface optimization, (b) new FOM definitions to capture metasurface reflectivity and other spectral properties, and (c) several user-defined features to enable more degrees of freedom for the optimizable geometry.

Additional modifications were made to the FOM specification, shown in Figure S3b, to capture metasurface reflection through a plane in free-space (and absorption by extension) as an optimization metric instead of modematching to a fundamental TE or TM mode in the original LumOpt implementation. Lastly, a number of new user-defined features were incorporated to enable more degrees of freedom for the optimizable geometry (shown in Figure S3c). These new features include: multi-axis optimization for both polarization-dependent and polarization-independent designs, and the ability to simultaneously optimize multiple objects to facilitate supercell structures.

**Adjoint Optimization Starting Design Analysis**

Here we identify the general boundaries of where adjoint optimization may fail to find adequate improvements to the initial structure. As shown in Figure S4, we used starting designs in the shape of cross-shaped resonators of various lengths and applied an optimization target of 7 μm. Starting cross lengths ranged from 2.2 to 2.8 μm in 0.2 μm steps, and the corresponding optimization results are shown in S4a to S4d, respectively. From these results, we observe that the starting cross length closest to the sought target (2.2 μm) yielded the highest final FOM (0.93) in the least number of iterations (8 iterations), while the final FOM decreased and the number of iterations increased as the starting structure gradually drifted further (up to 2.6 μm). Additionally, at the 2.8 μm cross length, we note that the optimization failed by finding negligible improvements to the starting structure ($\Delta$FOM=0.0025). Thus, based on our distance analysis, we estimate that the failure boundary was approximately 600 nm away from the target structure, which corresponded to an absorption peak of 1.5 μm from the target wavelength. Furthermore, the results here reinforce the importance of having a good starting design, since a design that is closer to the sought target can reduce the number of optimization iterations or total computation time (as demonstrated in the main text).

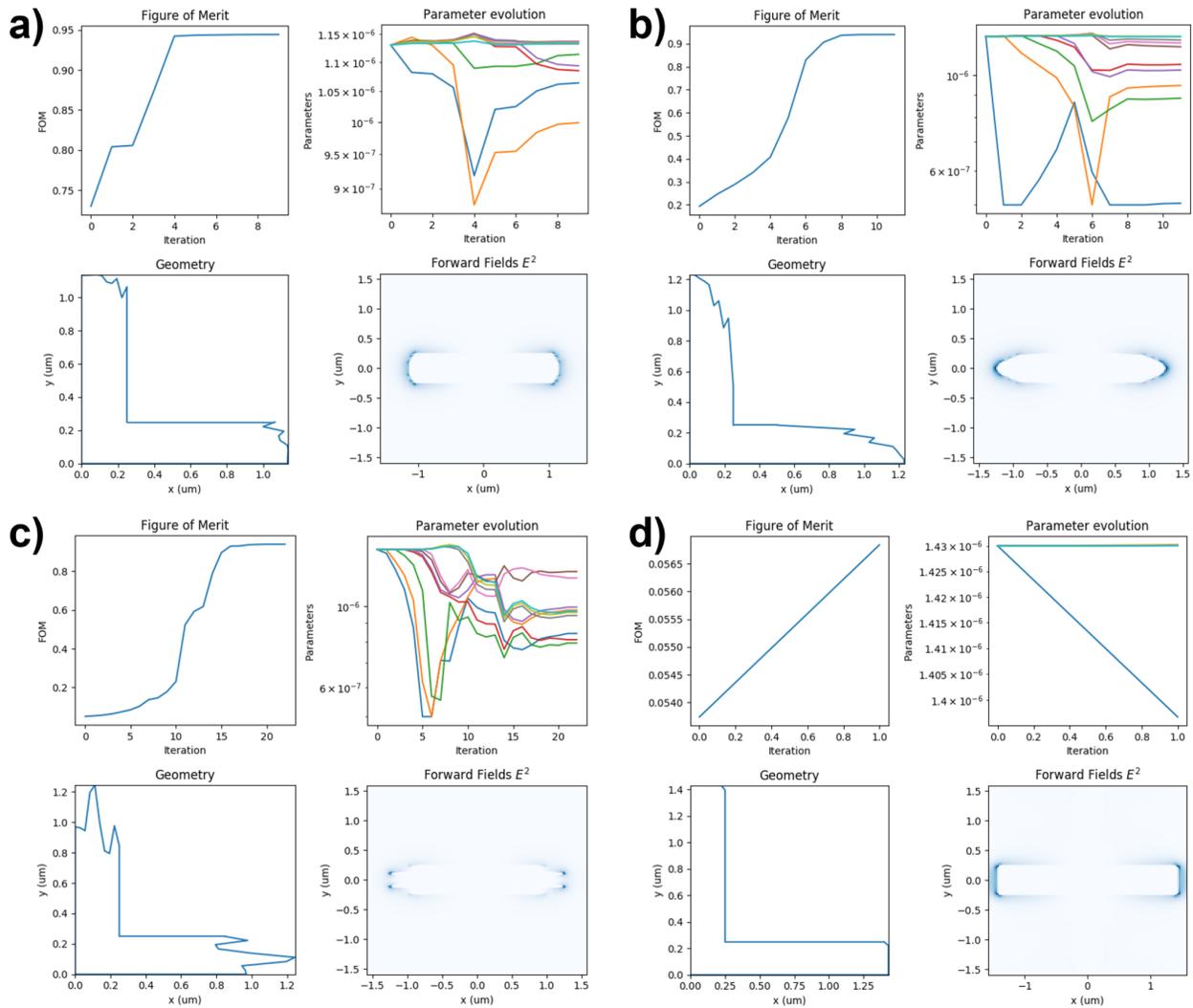

**Figure S4.** Adjoint optimization starting design distance analysis. Starting designs are cross lengths of (a) 2.2, (b) 2.4, (c) 2.6, and (d) 2.8 μm. Optimization target is 7 μm. Shown in each subplot are the FOM evolution, parameter evolution, and final geometry coordinates and mid-plane electric field profile. Computation time (number of iterations) increases as the starting design drifts further away from the sought target, until the optimization fails to find significant improvements to the FOM (here, at 2.8 μm).

For comparison, in Figure S5, we show the adjoint optimization results from DeepAdjoint for the same 7 μm target. Here, we observe that DeepAdjoint reduced optimization iterations by more than 50% (down to 2 iterations) and achieved a higher FOM (0.98). On a distributed high-performance computing cluster with four dedicated compute nodes, where a node has a minimum of four 64-bit Intel Xeon or AMD Opteron CPU cores and 8 GB memory, the DeepAdjoint optimizations corresponded to approximately 1 hour of total computation time.

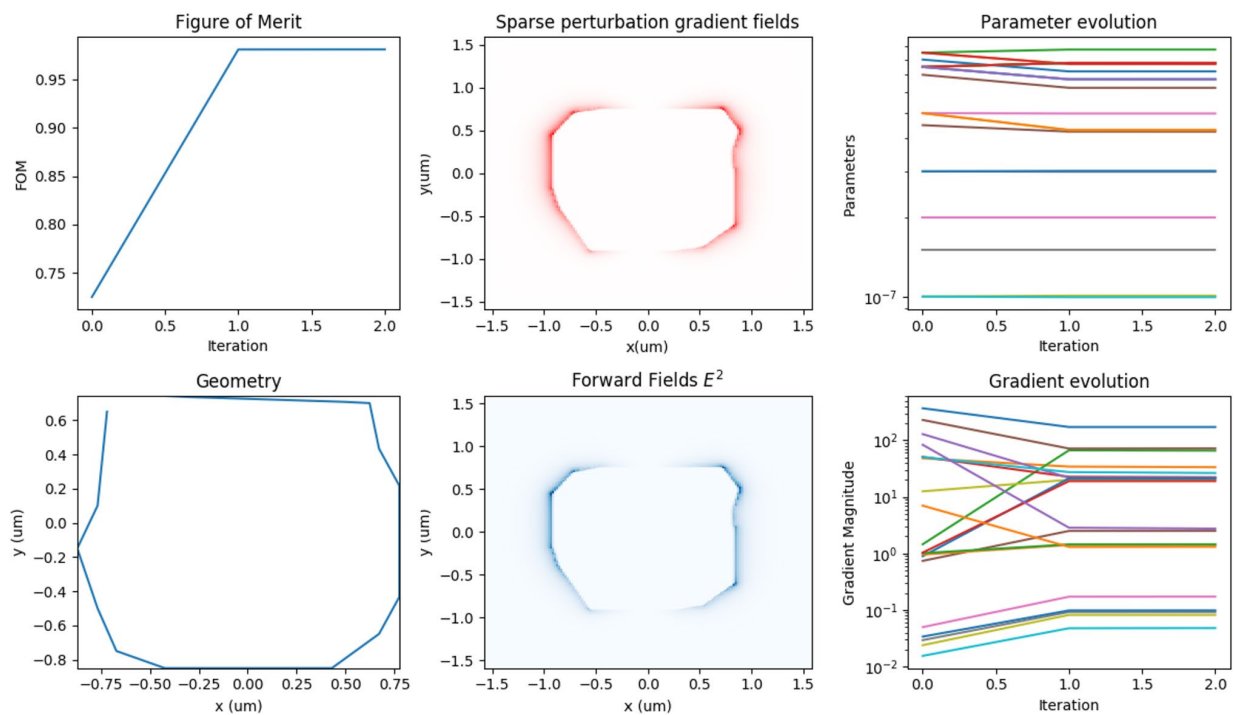

**Figure S5.** Device optimization using DeepAdjoint. Optimization target is 7 μm. In comparison to Figure S4, the DeepAdjoint optimization achieves higher performance in fewer iterations.

**Multi-Objective Metasurface Design (Increased Unit Cells)**

To demonstrate our DeepAdjoint approach for multi-objective designs with even greater complexity, we conducted an additional optimization with four unit cells and presented the results in Figure S6 below. First, we passed four Lorentzian-shaped spectra into the GAN (blue lines), which possessed absorption peaks at 6, 6.5, 7, and 7.5 μm. We then used the generated supercell (orange lines) as the starting point for adjoint optimization. In comparison to the initial structure, the optimized supercell yielded a 47% improvement in the FOM (yellow lines), thus validating the applicability of our approach to designs with greater complexity and number of optimizable elements. We note that the FOM used here is the total absorption averaged across all optimization target wavelengths (red dashed lines).

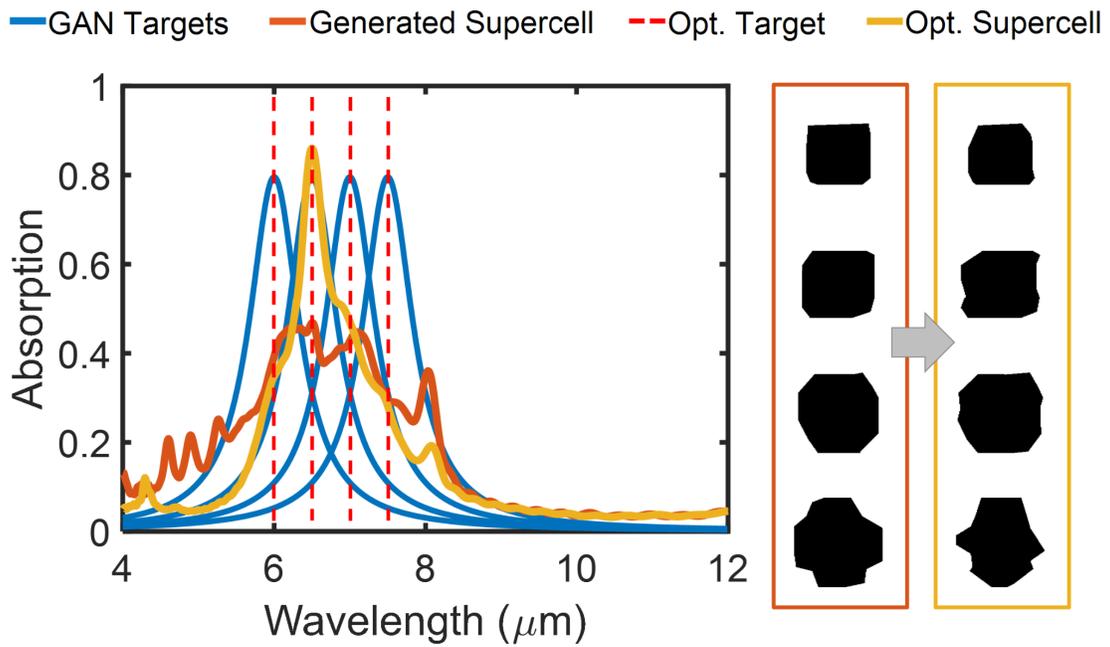

**Figure S6.** Multi-objective metasurface design with four unit cell elements. Target spectra (blue lines) are passed into the generative model (GAN) to produce starting designs (orange lines) for adjoint optimization (red dashed lines). Optimized designs (yellow lines) exhibit up to 47% performance enhancements in comparison to GAN-generated supercell designs (shown in the inset images).